\documentclass[twoside,reqno]{HERON}
\usepackage{epsfig,cite,colordvi}
\usepackage{graphicx}
\usepackage{url}
\usepackage{amssymb,amsmath,amscd,epsf}
\usepackage{times}
\usepackage{makeidx}
\usepackage{color}
\usepackage{hyperref}
\hypersetup{
	colorlinks   = true, 
	urlcolor     = blue, 
	linkcolor    = blue, 
	citecolor   = red 
}
\begin{document}
\title{Collective motion in  triaxial nuclei within minimal length concept }
\runningheads{Preparation of Papers for Heron Press Science Series
Books}{M. Chabab, A. El Batoul, A. Lahbas and M. Oulne}
\begin{start}
\coauthor{M. Chabab}{}, \author{A. El Batoul}{}\footnote{\textcolor{blue}{ elbatoul.abdelwahed@edu.uca.ma}}, \coauthor{A. Lahbas}{}, \coauthor{M. Oulne}{}

\address{High Energy Physics and Astrophysics Laboratory, Faculty of Sciences 
Semlalia, Cadi Ayyad University, P. O. B. 2390, Marrakesh 40000, Morocco.}{}
\begin{Abstract}
 The concept of minimal length, inspired by Heisenberg algebra, is applied to  the geometrical collective Bohr- Mottelson model (BMM) of nuclei. With the deformed canonical commutation relation and the Pauli-Podolsky prescription, we have derived the quantized Hamiltonian operator for  triaxial nuclei as we have previously  done for axial  prolate $\gamma$-rigid ones \textbf{\textcolor{blue}{(M. Chabab et al., Phys. Lett. B 758 (2016) 212-218)}}.
By considering an infinite square well like potential in $\beta$ collective shape variable, the eigenvalues of the Hamiltonian are obtained in terms of zeros of Bessel functions of irrational order with an explicit dependence on the minimal length parameter. Moreover, the associated symmetry with the model that we have constructed here can be considered as a new quasi-dynamical critical point symmetries (CPSs) in nuclear structure.
The theoretical results indicate a dramatic contribution (Low effect) of the minimal length to energy levels for lower values of the angular momentum and  regular (significant effect) for higher values. In fact, these features show that the minimal length scenario which is useful in recognizing the properties of real  deformed nuclei having high spins  or  strong rotations.
Finally, numerical calculations are performed for some nuclei such as  ${}^{124,128,130}$Xe and ${}^{114}$Pd revealing a qualitative agreement with the experimental data. 
\end{Abstract}
\end{start}
\section{Introduction}
The collective model of Bohr and Mottelson\cite{b1,b2} was designed to describe the collective low energy states of the nucleus in terms of rotations and vibrations of its ground state shape, which is parameterized by $\beta$ and $\gamma$ variables defining the deviation from sphericity and axiallity, respectively. Recently, considerable attempts have been done for several potentials to achieve analytical solutions of Bohr Hamiltonian, either in the usual case where the mass parameter is assumed to be a constant\cite{b3,b4,b5,b6,b7,b8} or in the context of deformation dependent mass formalism \cite{b9,b10,b11,b12}. Moreover, a great interest for solutions of this model has been revived with the proposal of E(5)\cite{b13} and X(5)\cite{b14} symmetries, which describe the critical points of the shape phase transitions between spherical and $\gamma$-unstable shapes and, from spherical to axial symmetric deformed shapes, respectively. The E(5) symmetry is a $\gamma$-independent exact solution of the Bohr Hamiltonian, while the X(5) symmetry is an approximate solution for $\gamma=0$. In addition, on the basis of the X(5) symmetry, a prolate $\gamma$-rigid version called X(3)\cite{b15} has been developed which is parameter independent and where the infinite square well potential has been used. Such symmetry has motivated the issue of the X(3)-$\beta^{2n}(n=1,2,3)$ models\cite{b16,b17} with a harmonic oscillator type for the $\beta$ potential. Recently, improved versions of the standard X(3) and X(5) symmetries, called X(3)-ML and X(5)-ML, have been developed with the introduction of the concept of minimal length\cite{b18}. In addition to axial shape, triaxial shapes in nuclei have also been inspected, since the introduction of the rigid triaxial rotor\cite{b19,b20}, despite the fact that very few candidates have been found experimentally\cite{b21,b22}. Furthermore, the softness model needs to be incorporated if a more complete description of the low-energy structure is considered. Soft triaxial nuclei have been studied, in this context, by making use of different types of potentials\cite{b23,b24,b25,b26}. 
Motivated by these considerations, other models such as Z(4)-sextic model, using Davydov-Chaban Hamiltonian for deformed nuclei have also been studied\cite{b27}. Therefore, our aim in the present work is to use the  minimal length formalism, as in Ref\cite{b18,b28,b29},  but this time in  the framework  of Davydov-Chaban model\cite{b30}. The next sections include a presentation of the Z(4) model with  minimal length scenario, its numerical results and  brief discussion for energy spectrum of some triaxial $\gamma$-rigid nuclei, while in the last section is devoted  to the main conclusions.
\subsection{Z(4) model with  minimal length: Z(4)-ML}
It is known that the nuclear excitations are determined by quadrupole vibrational-rotational degrees of freedom which can be treated simultaneously by considering, generally five quadrupole collective coordinates that describe the surface of a deformed nucleus. However, to separate rotational and vibrational motion, these coordinates are usually parametrized by two deformation parameters $\beta$ and $\gamma$ and three Euler angles $(\phi, \theta,\psi)$, which obviously define the orientation of the intrinsic principal axes in the laboratory frame. So, by imposing a $\gamma$-rigidity to the quadrupole collective motion, described in the  frame of the Bohr-Mottelson model\cite{b1,b2}, the corresponding eigenvalue problem  reduce to that of the Davydov-Chaban hamiltonian\cite{b30}:
\begin{equation}
H_{DC}=-\frac{\hbar^2}{2B_m}\left[\frac{1}{\beta^3}\frac{\partial}{\partial\beta}\beta^3\frac{\partial}{\partial\beta}-\frac{1}{4\beta^2}\sum_{k=1}^{3}\frac{Q_{k}^2}{\sin^2(\gamma-\frac{2\pi}{3}k)}\right]+ V(\beta),
\end{equation}
where $Q_k (k = 1, 2, 3)$ are the components of angular momentum and $B_m$ is the mass parameter.
By considering that the minimal length concept acting  only on the degree of freedom $\beta$ in accordance with the general rules of quantum mechanics, in the case of a curvilinear coordinates system, we obtain the Hamiltonian operator of the nucleus in the form :
\begin{align}
\hat{H}_{DC}^{ML}=-\frac{\hbar^2}{2B_m}\Delta +\frac{\alpha\hbar^4}{B_m}\Delta^2+ V(\beta),
\label{E15}
\end{align}
with
\begin{align}
\Delta = \Biggl[\frac{1}{\beta^3}
\frac{\partial}{\partial\beta}\beta^3\frac{\partial}{\partial\beta} -
\frac{1}{4\beta^2}\sum_{k=1}^{3}\frac{Q_{k}^2}{\sin^2(\gamma-\frac{2\pi}{3}k)}\Biggr] =-\frac{2B_m}{\hbar^2}\left( H_{DC}-V(\beta)\right).
\label{E16}
\end{align}
Here, $\gamma$ is treated as a parameter and not a variable. When $\gamma=30^\circ$, two moments of inertia in the intrinsic reference frame become equal, then the rotational term reads
\begin{equation}
\sum_{k=1}^{3}\frac{Q_{k}^2}{\sin^2(\gamma-\frac{2\pi}{3}k)}=\left(4\hat{Q}^2-3\hat{Q_1}^2\right).
\end{equation}
In order to determine the eigenfunctions and eigenvalues of the operator $\hat{H}_{DC}^{ML}$, we put 
\begin{equation}
\Psi(\beta,\Omega)=\left[1+2\alpha\hbar^2\Delta\right]F_{n_{\beta}}(\beta)\, Y_{\mu,\omega}^{L}(\Omega),
\end{equation}
where $\Omega$ denotes the rotation Euler angles $(\theta_1,\theta_2,\theta_3)$.
Therefore, we obtain the following equation in the variable $\beta$:
\begin{equation}
\Biggl[\frac{1}{\beta^3}
\frac{d}{d\beta}\beta^3\frac{d}{d\beta}
-\frac{W_{L,\omega}}{4\beta^2} + \frac{2B_m}{\hbar^2}\left(\frac{E-V(\beta)}{\left(1+4B_m\alpha\left(E-V(\beta)\right)\right)}\right)\Biggr]F_{n_{\beta}}(\beta) = 0. 
\label{E23}
\end{equation}
where $n_{\beta}$ is radial quantum number and  $W_{L,\omega}=\left(4\hat{Q}^2-3\hat{Q_1}^2\right)Y_{L,\omega}(\Omega)=4L(L+1)-3\omega^2$  is found by using the following symmetrized wave function,
\begin{equation}
Y_{\mu,\omega}^{L}(\Omega)=
\sqrt{\frac{2L+1}{16\pi^2(1+\delta_{\omega,0})}}\,
\Bigl[{\cal D}^{(L)}_{\mu,\omega}(\theta_i) 
+ (-1)^L {\cal D}^{(L)}_{\mu,-\omega}(\theta_i)\Bigr].
\end{equation}
Here, ${\cal D}(\theta_i)$ represent the Wigner functions of the Euler angles. The eigenvalues of the angular momentum in the intrinsic frame are given by $L$, while the projections of the angular momentum on the laboratory-fixed $\hat{z}$-axis and on the body-fixed $\hat{x}'$-axis are denoted $\mu$ and $\omega$, respectively.
It should be noted that in the limit $\alpha \rightarrow 0$, Eq. (\ref{E23}) reduces to the ordinary  Schr\"{o}dinger equation corresponding to the Hamiltonian $H_{DC}$ in Ref.\cite{b31} .
\subsubsection{Analytical solution for $\beta$ part: }
Here, we consider the above mentioned equation in the  previous section,  with an infinite square-well potential which is an important model for studying the  critical point symmetry in nuclei. The infinite square well potential defined by:
\begin{align}
V(\beta) = \left\{ \begin{array}{lcl} 0, && \rm{if}\ \beta\leq\beta_{\omega} \\
\infty, &&  \rm{if}\ \beta > \beta_{\omega} \end{array} \right. ,
\label{E25}
\end{align}
where $\beta_{\omega}$ indicates the width of the well. The peculiarity of this potential resides in the fact that it admits an infinite number of minimums and moreover, it obviously cannot have unbound states; all possible energies will therefore be quantified. The differential  "radial" Eq. (\ref{E23}) is exactly soluble in this case. Using the transformation of the wave function $F(\beta)=\beta^{-1}f(\beta)$, Eq. (\ref{E23}) becomes a differential Bessel equation
\begin{equation}
\Biggl[\frac{d^2}{d\beta^2} + \frac{1}{\beta}\frac{d}{d\beta}
+ \Biggl(\bar{k}-\frac{v^2}{\beta^2}\Biggr)\Biggr]f_{n_{\beta}}(\beta) = 0,
\label{E26}
\end{equation}
where we have introduced the following parametrization
\begin{equation}
\bar{k}=\frac{2B_m}{\hbar^2}\cdot\frac{E}{(1+4B_m\alpha E)}, \	v=\left[1+\frac{W_{L,\omega}}{4}\right]^{1/2}=\left[1+L(L+1)-\frac{3}{4}\omega^2\right]^{1/2}.
\label{E27}
\end{equation}
Then, the boundary condition $ f_{n_{\beta}}(\beta_{\omega})=0$  determines the energy spectrum for the $\beta$ degree of freedom  of the system,
\begin{equation}
E_{s,L} = \frac{\hbar^2}{2B_m}\times\frac{\bar{k}_{s,\eta}^2}{1-2\hbar^2\alpha\bar{k}_{s,\eta}^2},\ \ \bar{k}_{s,v}= \frac{x_{s,v}}{\beta_{\omega}},
\label{E28}
\end{equation}
and the corresponding eigenfunctions, which are finite at $\beta=0$, are then given by
\begin{equation}
F_{n_{\beta}}(\beta)=F_{sL}(\beta) = N_{s,L}\,\beta^{-1}
J_{\eta}(\bar{k}_{s,\eta}\beta),  \ s=n_{\beta}+1 .
\label{E29}
\end{equation}
with $ \bar{k}_{s,v}=x_{s,v}/\beta_{\omega}$ and $x_{s,v}$ is the $s$-th zero of the Bessel function of the first kind
$J_{v}(\bar{k}_{s,v}\beta_{\omega})$. $N_{s,L}$ is a normalization constant .
From a theoretical point of view, instead
of the parameter $\omega$, it is customary to use the wobbling quantum number $n_{\omega}=L-\omega$ for the  energy spectrum. In  the ground and  $\beta$ bands this quantum number is $n_{\omega}=0$, while for the $\gamma$-band it takes the values $n_{\omega}=1$ for $L$ odd and $n_{\omega}=2$ for $L$ even.
From the definition of the wobbling quantum number, the eigenvalue of the system is written as
\begin{equation}
E_{s,L,n_{\omega}}^{(\alpha,\beta_{\omega})} = \frac{\hbar^2}{2B_m}\times\frac{\left(\frac{x_{s,v}}{\beta_{\omega}}\right)^2}{1-2\hbar^2\alpha\left(\frac{x_{s,v}}{\beta_{\omega}}\right)^2}.
\label{E30}
\end{equation}
Since $\alpha$ is a very small parameter, we can develop this formula to the first order around $\alpha = 0 $, hence
\begin{equation}
E_{s,L,n_{\omega}}^{(\alpha,\beta_{\omega})} \simeq \frac{\hbar^2}{2B_m}\left(\frac{x_{s,v}}{\beta_{\omega}}\right)^2\left[1+2\hbar^2\alpha\left(\frac{x_{s,v}}{\beta_{\omega}}\right)^2\right],
\label{E31}
\end{equation}
with
\begin{equation}
 v=\frac{\sqrt{L(L+4)+3n_{\omega}\left(2L-n_{\omega}\right)+4}}{2}.
 \label{E32}
\end{equation}
It should be noticed here, that our expressions for the energy spectrum (\ref{E31}) reproduce exactly the results found in Ref.\cite{b31} in the $\alpha\rightarrow 0$ limit. For our  subsequent calculations, we define the energy
ratios as
\begin{equation}
R\left(s,L,n_{\omega}\right)=\frac{E_{s,L,n_{\omega}}^{(\alpha,\beta_{\omega})}-E_{1,0,0}^{(\alpha,\beta_{\omega})}}{E_{1,2,0}^{(\alpha,\beta_{\omega})}-E_{1,0,0}^{(\alpha,\beta_{\omega})}}.
\label{E33}
\end{equation}
\section{Results and Discussion}
\begin{figure}[h]
	\centering
	\includegraphics[height=24mm]{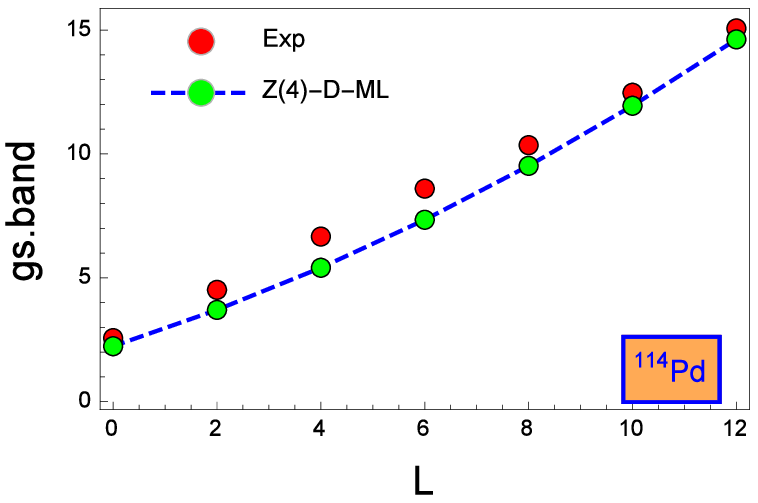}
	\includegraphics[height=24mm]{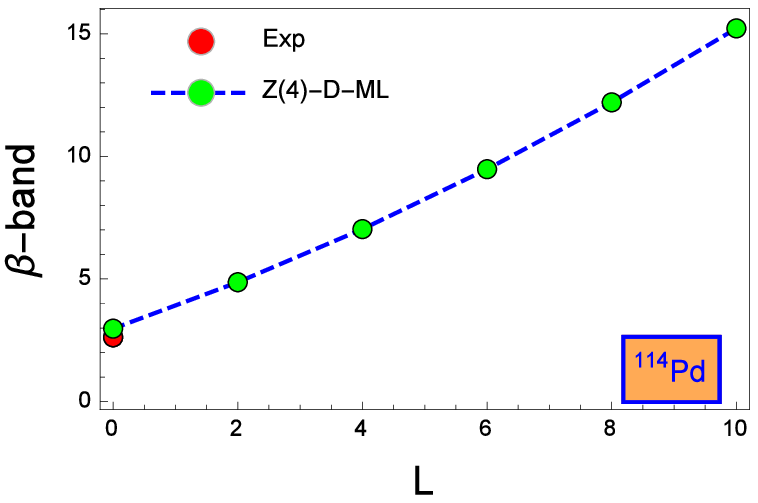}
	\includegraphics[height=24mm]{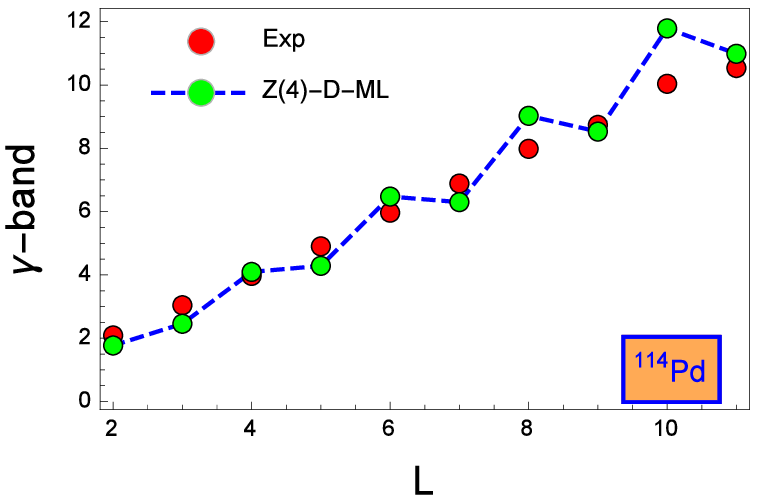}
	\includegraphics[height=24mm]{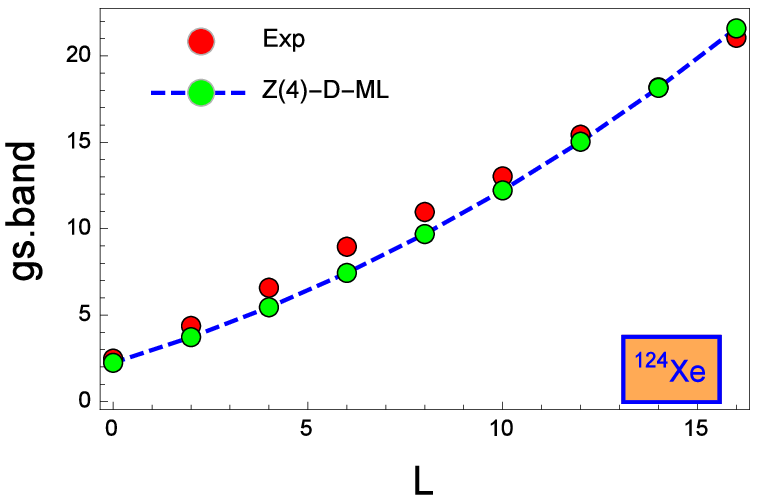}
	\includegraphics[height=24mm]{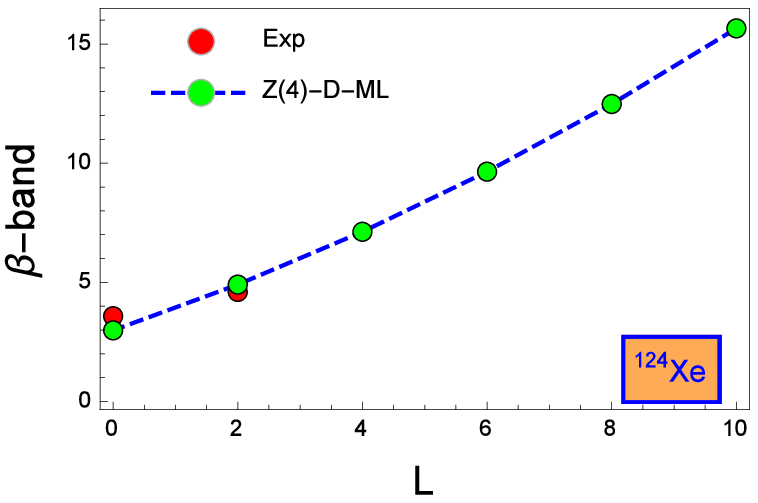}
	\includegraphics[height=24mm]{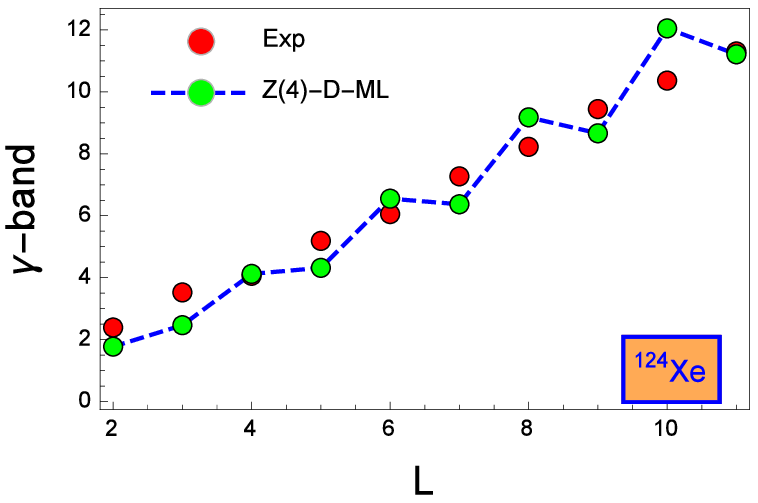}
	\includegraphics[height=24mm]{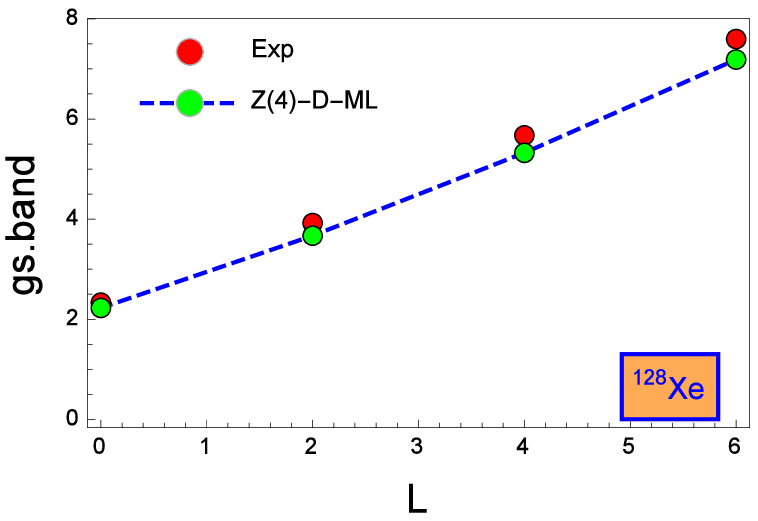}
	\includegraphics[height=24mm]{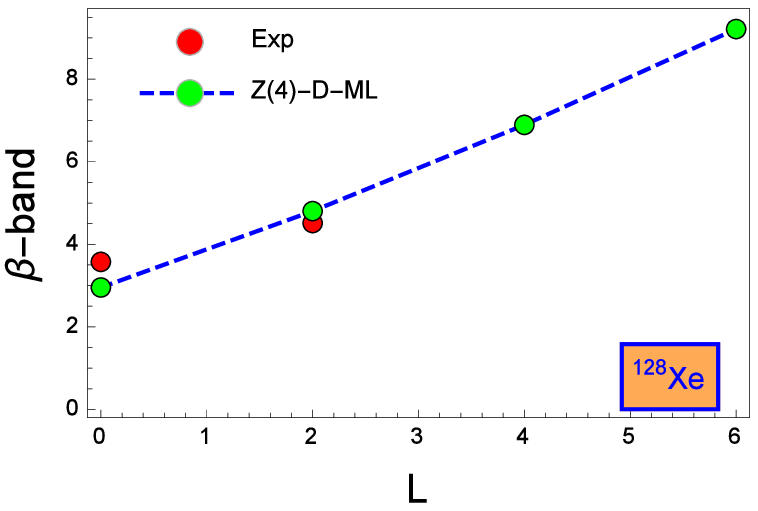}
	\includegraphics[height=24mm]{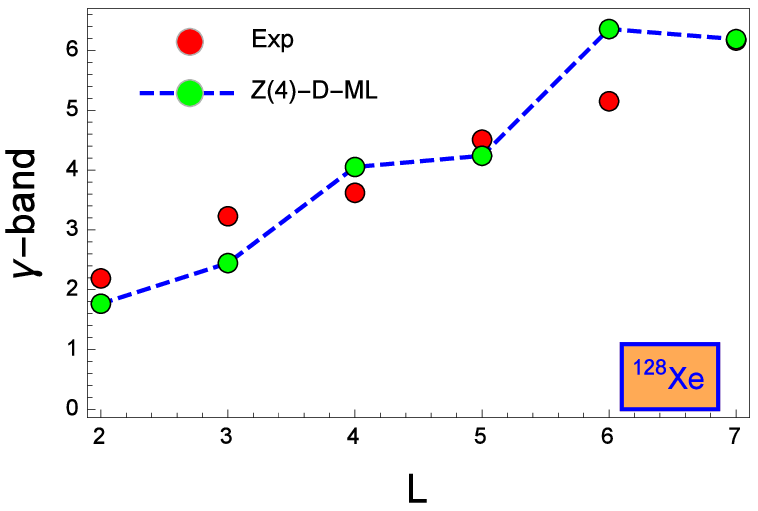}
	\includegraphics[height=24mm]{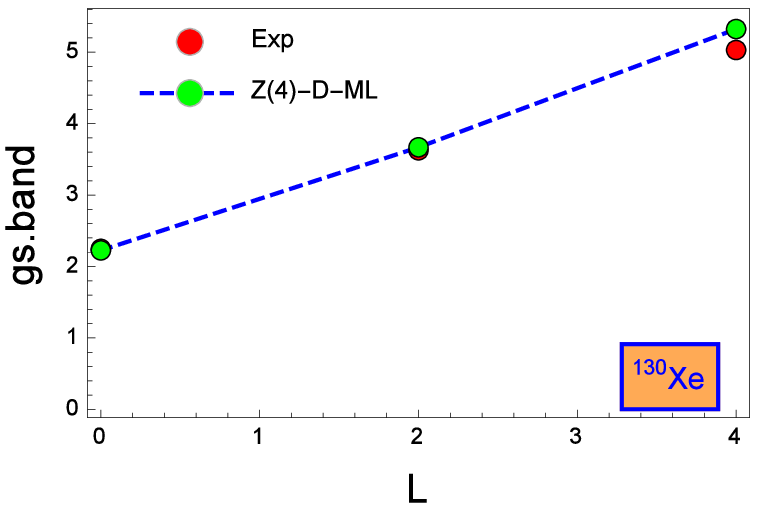}
	\includegraphics[height=24mm]{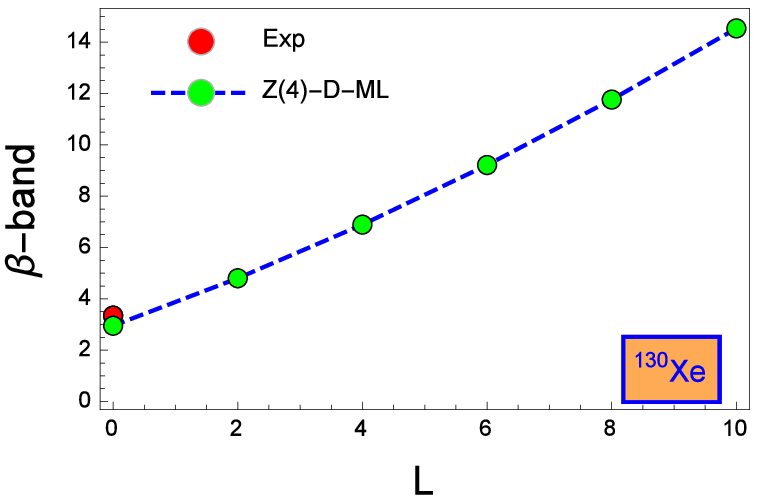}
	\includegraphics[height=24mm]{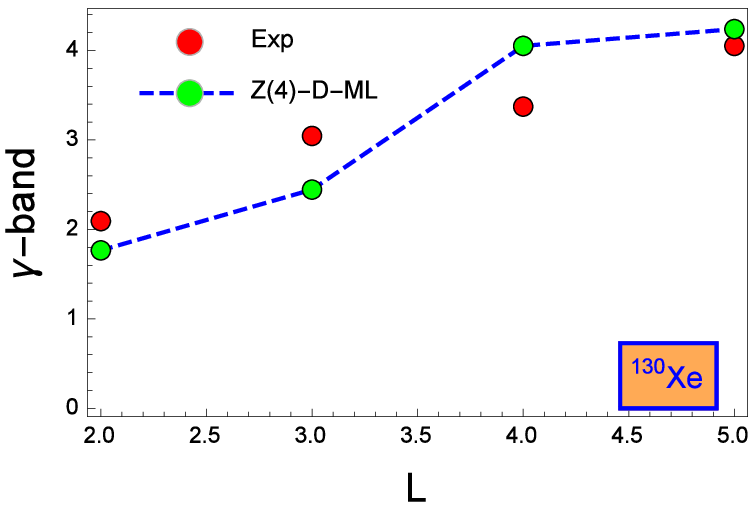}
	\caption{Comparison of the Z(4)-ML predictions for (normalized) energy levels  to experimental data\cite{b32} for ${}^{124,128,130}$Xe and ${}^{114}$Pd.}
	\label{Fig1}
\end{figure}

After presenting the theoretical framework of our model, we have now to perform the model analysis by focusing on its energy structure.So, before proceeding with any calculations of the energy spectra  for the triaxial nuclei, which  have a $\gamma$ rigidity of $30^\circ$, we briefly recall  a few interesting low-lying bands  which are classified by the quantum numbers $n_{\beta}$ ($s=n_{\beta}+1$), $n_{\gamma}$, and $n_{\omega}$, namely:
\begin{itemize}
 \item The ground state band (gsb) with $n_{\beta}=0$, $n_{\gamma}=0$, $n_{\omega}=0$,
 \item The $\gamma$ band composed by the even $L$ levels with $n_{\beta}=0$, $n_{\gamma}=0$, $n_{\omega}=2$ and the odd $L$ levels with $n_{\beta}=0$, $n_{\gamma}=0$, $n_{\omega}=1$,
 \item The $\beta$ band with $n_{\beta}=1$, $n_{\gamma}=0$, $n_{\omega}=0$.
\end{itemize}
It should be noted here, that  the ground, the $\beta$ and the $\gamma$ bands contain the rotational, the $\beta$ and $\gamma$ vibrational structures respectively.\\ The energy spectra of our model is predicted with two independent parameters $\beta_{\omega}$ and $\alpha$, whose values are obtained, via the least squares method, from fits to the experimental data.
The proposed model, called Z(4)-ML, is adequate for the description of $\gamma$-rigid nuclei having a triaxiallity close to $\gamma=30^\circ$. Moreover, the infinite square well potential allows the study of different $\beta$ deformations as in the pure model Z(4). Besides, to test the validity of the model, four nuclei (isotopes ${}^{124,128,130}$Xe and isotope ${}^{114}$Pd) are chosen as good  candidates  for triaxial nuclei . Moreover, the  parameters $\beta_{\omega}$ and $\alpha$ for each nucleus are obtained by fitting  their experimental energy spectrum comprising ground, $\beta$ and $\gamma$ bands with the energy formula (\ref{E33}), both being normalized to the corresponding energy of the first excited state. The mentioned nuclei were thus found to have the smallest deviations from the experimental data, evaluated by the quality measure $\sigma=\sqrt{\sum_{i}^{N}\left(E_i^{exp}-E_i^{Th}\right)^2/N}$, where $N$ is  the maximum number of considered levels.The values of the used free parameters in the calculations are listed in Table \ref{table1}. Moreover,the comparison between Z(4)-ML theoretical predictions and experimental data of selected candidates regarding energy levels is visualized schematically in Fig.\ref{Fig1}. The agreement with experiment is very good for the ground state band and $\gamma$ band, despite the fact that there is not much experimental data especially for the $\beta$ band of these studied nuclei. As a result, one concludes that the Z(4)- ML  is more suitable for describing the structural properties of nuclei having a structure in  vicinity to the Z(4) limit.
\begin{table}[h]
	\begin{center}
		{\linespread{2}
			\footnotesize
			\begin{tabular}{|c|c|c|c|}
				\hline
				Models&\multicolumn{2}{c|}{Z(4)-ML}	\\
				\hline
				Nucleus&$\alpha$&$\beta_{\omega}$\\
				\hline
				$ {}^{114}$Pd &0.00006302& 0.64712\\
				$ {}^{124}$Xe &0.00010959& 0.66447\\
				$ {}^{128}$Xe &0.00000012& 0.89820\\
				$ {}^{130}$Xe &0.00000091& 0.89474\\
				\hline
		\end{tabular}}
	\end{center}
	\caption{The values of the free parameters used in the calculations.}
\label{table1}
\end{table}

\section{Conclusions}
In this work, we have derived new solutions of the Bohr-Mottelson Hamiltonian in the triaxial $\gamma$-rigid regime within the minimal length formalism where the shape of the potential of the collective $\beta$-vibrations is assumed to be equal to an infinite square well as in the standard Z(4)  model. Howeover, improved version of the Z(4) symmetry denoted Z(4)-ML  is elaborated for the first time in order to describe the structural properties  of some  triaxial $\gamma$-rigid nuclei. Finally, through this work, one can conclude that the introduction of the minimal length concept in Z(4) allows one to enhance the numerical calculation precision of the energy spectrum of  some  triaxial $\gamma$ -rigid nuclei in comparison with the Z(4) model.
\section*{Acknowledgments}
A. El Batoul  would like to thank Prof. M. Gaidarov and Prof N. Minkov for their kind invitation to the international Workshop on nuclear physics \textcolor{blue}{"36-th INTERNATIONAL WORKSHOP ON NUCLEAR THEORY"} and for their hospitality. He is thankful to Prof. M. Minkov for enlightening discussions, especially those concerning the numerical  calculation of the parameters $\alpha$ and $\beta_{\omega} $. Also, he acknowledges the financial support (Type A) of Cadi Ayyad University.

\end{document}